\documentstyle[12pt]{article}
\newcommand{\plabel}{\label}
\begin{document}
\begin{titlepage}
\renewcommand{\thefootnote}{\fnsymbol{footnote}}

\hfill TUW--98--22  \\
\begin{center}
\vspace{1cm}

{\Large\bf  Effective action and Hawking radiation for dilaton
coupled scalars in two dimensions}
\vspace{1.0cm}
\vfill
\renewcommand{\baselinestretch}{1}

{\bf W.\ Kummer$^1$\footnotemark[1] and  
 D.V.\ Vassilevich$^{2}$\footnotemark[2]
\footnotemark[3]}
\vspace{7ex}

{$^1$Institut f\"ur
    Theoretische Physik \\ Technische Universit\"at Wien \\ Wiedner
    Hauptstr.  8--10, A-1040 Wien \\ Austria}
\vspace{2ex}

{$^2$Erwin Schr\"odinger International Institute for Mathematical 
Physics, \\ Boltzmanngasse 9, A-1090 Wien,\\ Austria\\[0.2cm]
$^2$Institut f\"ur Theoretische Physik,\\ Universit\"at Leipzig,\\ 
Augustusplatz 10, D-04109 Leipzig,\\ Germany}

\footnotetext[1]{e-mail: {\tt wkummer@tph.tuwien.ac.at}}
\footnotetext[2]{e-mail: 
{\tt Dmitri.Vassilevich@itp.uni-leipzig.de}}
\footnotetext[3]{On leave from Department of Theoretical Physics,
   St. Petersburg University, 198904 St. Petersburg, Russia}

\end{center}
\vfill

\begin{abstract}
The effective one-loop action for general dilaton theories with
arbitrary dilaton-dependent measure and nonminimal coupling to
scalar matter is computed.  As an application we determine the
Hawking flux to infinity from black holes in $D$-dimensions.  We
resolve the recently resurrected problem of an apparent negative
flux for nonminimally coupled scalars: For any $D \geq 4$ Black
Hole the complete flux turns out to be precisely the one of
minimal coupling.  This result is obtained from a
Christensen-Fulling type argument involving the
(non-)conservation of energy-momentum.  
It is compared with approaches using the effective action.
\end{abstract}
\end{titlepage}
\vfill

\section{Introduction}
The study of gravity models in two dimensions provides some 
important answers to the difficult questions posed by the quantization 
of gravity. Indeed, the restriction to one time and one space 
coordinate of Einstein gravity in four and higher dimensions, to 
a $D = 2$ dilaton theory (spherical reduction, SRG) \cite{thom84} 
represents a class of such models with immediate physical 
relevance.

The last decade has seen substantial progress in this field. 
Especially the use of a light cone gauge for the Cartan-variables 
\cite{KS92} which amounts to choosing an Eddington-Finkelstein 
gauge for the 2d metric has led to a much better and simpler 
understanding of the classical theory \cite{Gro92,KKL97} as well 
as at the quantum level \cite{KS92-2}. Most of these results have 
turned out to be less obvious or even not attainable in the 
traditional approach where a conformal gauge was used for the 2d 
metric \cite{Man91,deAl92}. 

In order to be able to attribute to 2d models of gravity a 
sufficient credibility as far as the application of their results 
to the genuine (higher dimensional) case is concerned, an obvious 
precondition is that what has been found already at $D \geq 4$ 
should be fully reproduced in $D = 2$, wherever such an overlap 
occurs and can be tested. Hawking radiation, one of the most 
interesting features of Black Hole (BH) physics, is a consequence of 
precisely this kind. It is known from calculations in $D = 4$ 
that the thermal radiation from a BH to infinity is 
related to the Hawking temperature at the horizon according to 
the laws of black body radiation \cite{CF77}. A central 
role is played by the scale anomaly of the energy momentum (EM) 
tensor. 

By spherical reduction from $D \geq 4$ the scalar field acquires
nonminimal coupling to the dilaton field.  That this may cause
complications in a fully twodimensional calculation of the
Hawking flux has been realized only relatively recently in the
pioneering paper by V.\ Mukhanov and collaborators
\cite{Mukhanov}.  As observed for the first time by these
authors, a naive adoption of the 4D approach due to the
dependence of the anomaly of the dilaton field leads to a
negative flux at infinity.  Therefore, these authors added to
their integrated effective (Polyakov) action a non local
Weyl-invariant term of the Coleman-Weinberg type which depends
on a renormalization scale.  Then the sign of the resulting
total flux was seen to become positive.

This question was taken up in a number of papers with mutually
contradicting results for the anomaly
\cite{Chiba,c-anom,KLVmpla,antiev,Ich}.  Using different
expressions for the effective action it was claimed
\cite{c-anom,antiev} that instabilities and ``anti-evaporation''
phenomena can occur.  Even the conformal anomaly itself became
the subject of a discussion
\cite{Chiba,c-anom,KLVmpla,Ich,KLVprd}.  In our opinion this
part of the problem has been settled with the enlightening paper
by Dowker \cite{Dowker} who confirmed the previous results
\cite{Mukhanov,Chiba,Ich} for SRG and our result \cite{KLVmpla}
for general dilaton models.  A recent summary of the different
approaches leading to negative Hawking flux and a general
framework for a solution of this problem has been proposed in
\cite{Balbinot}; another attempt to solve this problem can be
found in \cite{Russo}.

We believe that our present work for the first time provides a
complete but perhaps surprising answer to the question of $2d$ Hawking
radiation from nonminimally coupled scalar fields.  Our approach
is based upon a consistent use of $\zeta$-function
regularization, not only for the part of the effective action
determined by minimally coupled scalars, but also for the part
controlled by the dilaton field.

In Section 2 we recall the action of SRG in $D \geq 4$
dimensions.  We summarize our conventions for the solution of
this (BH) background part.  We also derive in a simple manner
the nonconservation relation for the EM tensor, valid for
arbitrary dilaton theories in any dimension for nonminimally
coupled matter.  The integrated effective action is determined
in Section 3.  Our result generalizes the Polyakov action
\cite{Polyakov} to arbitrary nonminimal dilaton coupling of
matter fields and to arbitrary dilaton dependent measure. 
Section 4 contains the application of the results of Sect.\ 3 to
the Hawking flux at infinity.  We first integrate the
nonconvervation equation for the EM tensor.  Here the input is
the 2d anomaly together with the result for the ``dilaton
anomaly'', not requiring the knowledge of the functionally
integrated action.  Subsequently we discuss how one may arrive
at this flux
from the EM tensor, obtained directly from
that action.

\section{Dilaton theory for spherical reduction}

\subsection{Spherically reduced action}

SRG in $D \geq 4$ is based upon the choice of the D-dimensional 
metric

\begin{equation}
(ds)^2 =g_{\mu\nu}dx^\mu dx^\nu - e^{-\frac{4}{D-2} \phi} (d\Omega )^2
\ ,
\plabel{dline}
\end{equation}
where $d\Omega$ is the standard line element on $S^{D-2}$. The dilaton
field $\phi$ and $g_{\mu\nu} $ depend on the two first coordinates $x^\mu$ 
only. In terms of (\ref{dline}) the  $D$-dimensional 
Einstein-Hilbert Lagrangian reduces to 

\begin{equation}
{\cal L}_{SRG} = e^{-2\phi} \sqrt{-g} \left(
R+\frac{4(D-3)}{D-2} (\nabla \phi )^2 
-\frac{1}{4} (D-2)(D-3) e^{\frac{4}{D-2} \phi}
\right) \plabel{Lsrg}
\end{equation}
which is the particular case 
 $a=\frac{D-3}{D-2}$ and $B= -\frac{1}{4} (D-2)(D-3)$ 
of a more general dilatonic Lagrangian treated in \cite{KKL97} 
for general values of the parameter $0 \leq a \leq 1$ 

\begin{equation}
\plabel{ldil}
{\cal L}_g =  \sqrt{-g}e^{-2\phi}(R+4a(\nabla\phi)^2 +Be^{4(1-a)\phi}).
\end{equation}

The family of models of this type comprises all theories with one 
horizon, Minkowski asymptotics and (for $0 < a < 1$) with the 
same (null- and non-null incomplete) singularity as the 
Schwarzschild BH. The dilaton BH \cite{Man91} is contained as the 
limit $a = 1$ or $D \to \infty$. It has null-complete geodesics 
at the singularity \cite{KKL97}. 

The most convenient way to obtain the general solution for
(\ref{Lsrg}) or (\ref{ldil}) has been described in \cite{KKL97}. 
For our present purposes we need the background solution in
conformal gauge.  With the proper choice of the coordinates
\cite{KKL,LVA} which yields the Minkowski metric in the
asymptotic region it takes the form

\begin{eqnarray}
(ds)^2 & =& K(u) (d\tau^2 -dz^2) \qquad dU= K(u) dz 
\plabel{4}\\
 K(u) & = & 1 - \left( \frac{u_h}{u} \right)^{D-3} 
\plabel{5}\\
\phi (u)  & =& - \frac{D-2}{2} \log (\frac{2u}{D-2})
\plabel{6}
\end{eqnarray}
where $u_h$ is the value of $u$ at the horizon (defined by the
equation $K(u_h)=0$), the asymptotic region corresponds to $u =
\infty$.  The explicit expression for $u_h$ is not needed for
our calculations, $u_h$ is proportional to the absolutely
conserved quantity $C$, which, in turn, is proportional to the
ADM mass of the BH \cite{LVA}.  For $D=4$ one obtains the usual
Schwarzschild solution.  In the solution (\ref{4})-(\ref{6}) we
slightly change the notation of \cite{LVA} ($U \to u, L(U) =
-K(u)$).  The line element (\ref{4}) in conformal light cone
coordinates reads

\begin{equation}
(ds)^2 = e^{2\rho}dx^+dx^- , \rho =\frac 12 \log (K (u) )\; ,
\plabel{7}
\end{equation}
where $x^\pm =\tau \pm z$.  The range for $z$ is $- \infty \leq z 
\leq + \infty$. 
Thus derivatives of light cone 
coordinates,  acting on functions of $u$ become

\begin{equation}
\partial_+ = -\partial_-=\frac 12 \partial_z =-\frac 12 
K(u)\partial_u\; .
\plabel{8}
\end{equation}
For completeness we also quote the Hawking temperature, computed 
from $K(u)$ as the surface gravity at the horizon:

\begin{equation}
T_H = \frac{1}{4\pi} K'(u)\vert_{u=u_h} = \frac{D-3}{4\pi u_h}\; .
\plabel{9}
\end{equation}

Reducing the action for massless matter $f$, coupled minimally in 
$D=4$ according to (\ref{dline}) leads to nonminimal coupling in the 
corresponding 2d Lagrangian 

\begin{equation}
{\cal L}_{(m)} = \frac{1}{2}\; e^{-2\phi}\; \sqrt{-g}\; 
g^{\mu\nu}\; (\partial_\mu f)\; (\partial_\nu f)\; .
\plabel{10}
\end{equation}

The  formalism in the following will be developed for a 
general dilaton factor $\exp (-2\varphi(\phi))$. 

Spherical reduction also affects the definition of the covariant
measure.  This is seen most directly from the path integral
whose diffeomorphism invariant definition at general $D$
requires a factor $\sqrt[4]{-g_{(D)}}$ where $g_{(D)}$ is the
determinant of the original $D$-dimensional metric \cite{Toms}. 
By Eq.\ (\ref{dline}) this yields a factor $e^{-\phi}$ so that
the scalar field $\tilde f$ redefined as $\tilde f = f\,
e^{-\phi}$ possesses a trivial measure.  Of course, such a
factor is nothing else but the inverse power of the radius
required for a proper inclusion of s-wave excitations.  Also
here we consider the more general case

\begin{equation}
\tilde f = f\, 
e^{-\varphi(\phi)}\;  
\plabel{11}
\end{equation}
and take the standard path integral measure for $\tilde f$.
Namely, we require $\int\, (d\tilde f)\, \sqrt[4]{-g}\,
 \exp (i\int \sqrt{-g}{\tilde f}^2)$
be a field independent (infinite) constant.

%%%%%%%%%%%%%%%%%%%%%%%%%%%%%%%%%%%%%%%%%%%%%%%%%%%%%%%%%%
\subsection{Nonconservation of the energy momentum tensor for 
dilaton coupled fields}

For nonminimal coupling of the scalars to the dilaton field the
conservation law for the EM tensor must be modified. 
Classically the matter field action is invariant under the
diffeomorphism transformations

\begin{eqnarray}
\delta g_{\mu\nu} & = & \nabla_\mu \xi_\nu + \nabla_\nu 
\xi_\mu\; , \nonumber \\
\delta \Phi & = & \xi^\nu \partial_\nu \Phi \plabel{diff}\; ,\\
\delta f & = & \xi^\nu \partial_\nu f\; , \nonumber
\end{eqnarray}
where $\Phi$ denotes either the dilaton field $\phi$ or any local
function thereof.  By applying the transformation (\ref{diff}) to 
the action $S_{(m)}$ we obtain on 
the mass-shell for the scalar field

\begin{equation}
\nabla^\mu T_{\mu\nu} =-(\partial_\nu \Phi )
\frac{1}{\sqrt{-g}} \frac{\delta S_{(m)}}{\delta \Phi} \; ,
\plabel{cons}
\end{equation}
where the EM tensor is defined as usual

\begin{equation}
\frac{\sqrt{-g}}2 T_{\mu\nu} =\frac{\delta S_{(m)}}{\delta 
g^{\mu\nu}}\; .
\plabel{emt}
\end{equation}
The symmetry (\ref{diff}) is retained also at the quantum level
when the scalar field is integrated out.  Thus Eq.\ (\ref{cons})
holds as well for the expectation values, i.e.\ for the
corresponding quantities computed from the one-loop effective
action $W$.  Recently the appearance of a nonconservation
equation (\ref{cons}) in $D = 2$ has been noted after reduction
from the $D = 4$ case \cite{Balbinot,Russo}.  As seen from the
simple derivation above such a relation holds in fact for any
generic dilaton theory in any dimension.  It could also be
interpreted as an ``extended'' conservation law, involving
$\delta W/\delta \Phi$ as part of an extended EM tensor.

\section{Effective action for dilaton theories}

Expressing the classical action related to Lagrangian (\ref{10}) in 
terms of the field $\tilde f$ according to Eq.\ (\ref{11}) yields the 
classical action 

\begin{equation}
\plabel{tact}
S = -\frac 12 \int \sqrt{-g}d^2x\ 
\tilde f A \tilde f \quad ,
\end{equation}
containing the differential operator 

\begin{equation}
A = -e^{-2\varphi +2\psi}g^{\mu\nu} (\nabla_\mu \nabla_\nu
+2(\psi_{,\mu}-\varphi_{,\mu})\nabla_\nu + 
(\nabla_\mu\nabla_\nu\psi) 
-2\varphi_{,\mu} \psi_{,\nu} +\psi_{,\mu} \psi_{,\nu} )\ . 
\plabel{A}
\end{equation}

The one loop effective action is obtained by the 
path integral for $\tilde f$ as  

\begin{equation}
W = \frac 12  \mbox{Tr} \ln A \quad .
\end{equation}
W depends on the metric, on $\varphi$ and $\psi$ which in the 
following all will be regarded as independent (background) 
fields. In Eq.\ (17) $W$ represents the Euclidean action. The 
path integral leading to that equation should be done with 
$\sqrt{-g} \to i\, \sqrt{g}$ in Eq.\ (\ref{tact}) to obtain the 
$\zeta$-function regularization method with elliptic differential 
operator $A$ after continuation to the Euclidean domain. This is 
implied in the following, although we retain Minkowski space 
notation. 

In  that regularization \cite{zeta1,zeta2} 
$W$ can be expressed in terms of the zeta function of the operator 
$A$:

\begin{equation}
W=-\frac 12 \zeta'_A(0), \qquad \zeta_A(s)={\rm Tr}(A^{-s})\; .
\plabel{zeta}
\end{equation}
Prime denotes differentiation with respect to $s$.

Evaluation of the ${\zeta'}_A (0)$ in general is quite a tedious
task.  For the case of a generic operator $A$ no analytic
formulas are available.  Fortunately, as will be shown below, in
the particular case of Eq..\ (\ref{A}) variations of the
$\zeta'_A(0)$ with respect to the dilaton field and to the scale
transformation of the metric can be reduced to known heat kernel
coefficients for certain second order operators.

First, we repeat our derivation \cite{KLVmpla} of the 
the trace of the EM tensor.
The variation of the zeta function with respect to a certain 
parameter or field is related to the one of the 
operator A as \cite{Atiyah,EKP}
\begin{equation}
\delta \zeta_A(s) =-s{\rm Tr}((\delta A) A^{-1-s})\ .
\plabel{var-z}
\end{equation}
An infinitesimal conformal transformation 
$\delta g_{\mu\nu} = \delta k(x) g_{\mu\nu}$ produces the 
trace of the (effective) EM tensor
\begin{equation}
\delta W=\frac 12 \int d^2x \sqrt{-g}\, \delta g^{\mu\nu}T_{\mu\nu}
=-\frac 12 \int d^2x \sqrt{-g}\, \delta\, k(x)\,T_\mu^\mu (x)\; .
\plabel{T}
\end{equation}
Due to the multiplicative transformation property $\delta A =
-\delta k A$ of Eq.\ (16) (valid in $D = 2$ only) powers of $A$
in Eq.\ (\ref{var-z}) recombine to $A^{-s}$.  With the
definition of a generalized $\zeta$-function \cite{Gilkey}

\begin{equation}
\zeta (s|\delta k,A)={\rm Tr}(\delta kA^{-s})
\plabel{varW}
\end{equation}
the variation in Eq.\ (\ref{T}) can be identified with 
\begin{equation}
\plabel{TX}
\delta W=-\frac 12 \zeta (0|\delta k,A) \quad .
\end{equation}
Combining Eqs.\ (\ref{TX}) and (\ref{T})  one obtains 
\begin{equation}
\zeta (0|\delta k,A)=\int d^2x \sqrt{-g} \delta 
k(x)T_\mu^\mu (x)\; .
\plabel{T2}
\end{equation}

By a Mellin transformation one can show that 
$\zeta (0|\delta k,A)=a_1(\delta k,A)$ \cite{Gilkey}, where $a_1$ is
defined as a coefficient in a small $t$ 
asymptotic expansion of the heat kernel:
\begin{equation}
{\rm Tr}(F\exp (-At)) =\sum_n a_n (F,A)t^{n-1}\; .
\plabel{hk}
\end{equation}
To evaluate $a_1$ according to  \cite{Gilkey} 
we rewrite  $A$ as ($\hat\nabla_\mu$ refers to the metric $\hat 
g_{\mu\nu}$) 
\begin{equation}
A=-(\hat g^{\mu\nu}D_\mu D_\nu +E) ,\qquad 
E=\hat g^{\mu\nu}(-\varphi_{,\mu}\varphi_{,\nu}
+ \hat\nabla_\mu \hat\nabla_\nu\, \varphi ) \; ,
\plabel{newA}
\end{equation}
where $\hat g^{\mu\nu}=e^{-2\varphi +2\psi}g^{\mu\nu}$,
$D_\mu = \hat\nabla_\mu +\omega_\mu$, $\omega_\mu =\psi_{,\mu}-
\varphi_{,\mu}$. 
Then for $a_1$ follows \cite{Gilkey}
\begin{equation}
a_1 (\delta k, A)=\frac 1{24\pi} {\rm tr}
\int d^2x\sqrt{-\hat g} \delta k  
(\hat R+6E)\; .
\plabel{a1}
\end{equation}
$\rm tr$ denotes ordinary trace over all matrix indices (if
any).  In the present case this is a trivial operation. 
However, below we will need the heat kernel coefficient $a_1$
for a matrix operator where the more general formula (26) is
essential.  Returning to the initial metric and comparing with
Eq.\ (\ref{T}) the most general form of the `conformal anomaly'
for non-minimal coupling in $D=2$ is found to be \cite{KLVmpla}

\begin{equation}
T_\mu^\mu =\frac 1{24\pi} (R-6(\nabla \varphi )^2 +
4\Box \varphi +2\Box \psi )\; .
\plabel{T3}
\end{equation} 

The variation of the effective action with respect to $\varphi$
or $\psi$ does not exhibit the same multiplicative property as
the conformal variation, because after substituting in Eq.\
(\ref{var-z}) the variation of $A$ does not recombine to powers
of $A$.  Therefore, the heat kernel technique is not applicable
to the evaluation of Eq.\ (\ref{var-z}) as it stands.  However,
crucial simplifications occur after transition to flat space by
means of a conformal transformation.  In conformal gauge
$g_{\mu\nu}=e^{2\rho}\eta_{\mu\nu}$ with flat metric
$\eta_{\mu\nu}$ the identity

\begin{equation}
\frac{\delta W(\rho )}{\delta\varphi} =
\int_0^\rho d\sigma \frac{\delta^2 W(\sigma )}{\delta \sigma
\, \delta\varphi} +\frac{\delta W(0 )}{\delta\varphi}
\plabel{toflat}
\end{equation}
is obvious, with an analogous one for the variation with respect
to $\psi$.
The first term on the right hand side of Eq.\ (\ref{toflat})
can be expressed in terms of the conformal anomaly:
\begin{equation}
\frac{\delta W(\rho )}{\delta\varphi} =
-\int_0^\rho d\sigma\, \sqrt{-g}\, 
\frac{\delta (\sqrt{g} T_\mu^\mu (\sigma ))}{ \delta\varphi} 
+\frac{\delta W(0 )}{\delta\varphi}\; .
\plabel{toflat2}
\end{equation}
To evaluate the second term, which at $\rho = 0$ represents the flat 
space contribution, we rewrite  $W(0)$  as
\begin{equation}
W(0)  = \frac 14 \log \int (d\vec{f})\; 
\exp \left( -\int d^2 x \sqrt{-\eta}\; \vec{f}\; {\bf 1}_2\, 
(A)\,\vec{f} \right)\; ,
\plabel{vecf}
\end{equation}
where we have doubled bosonic degrees of freedom by introducing
the two-component field $\vec{f}$. In flat space
the integral in the 
exponential in Eq.\ (\ref{vecf}) can be rewritten as

\begin{equation}
\int d^2 x \sqrt{-\eta}\; \vec{f}\; {\bf 1}_2\, 
(A)\,\vec{f} = \int  d^2 x \sqrt{-\eta}\, \vec{f}\, D\, 
D^{\dag}\, \vec{f}\; .
\plabel{DD}
\end{equation}
Here new differential operators in spinor space 
$D=i\gamma^\mu e^\psi \partial_\mu e^{-\varphi}$ and $ D^{\dag} = 
D(\psi \leftrightarrow - \varphi)$ have been introduced. 
Indeed, the right hand side of Eq.\ (\ref{DD}) is equal to

\begin{equation}
\int  d^2 x \sqrt{-\eta} \left[ \vec{f} \left(
A +2\gamma^5 \epsilon^{\mu\nu}e^{2(\psi -\varphi )}
\varphi_{,\mu}\psi_{,\nu} \right) \vec{f} +
\epsilon^{\mu\nu}e^{2(\psi -\varphi )} \varphi_{,\mu}
\partial_\nu (\vec{f}\gamma^5 \vec{f}) \right]
\end{equation}
which proves Eq.\ (\ref{DD}) 
after integration by parts. Therefore, in flat space
\begin{equation}
 2W (0) \; =\; \frac 12 \log \det (D\, D^{\dag} )
\plabel{ADD}
\end{equation}
holds. For the $\zeta$-function of the operator $D\, D^{\dag}$ we use
its representation in terms of an inverse Mellin transform
of the heat kernel
\begin{equation}
\zeta_{D\, D^{\dag}}\; (s) = \frac 1{\Gamma (s)} \int_0^\infty 
dt\, t^{s-1} {\rm Tr} \exp (-tD\, D^{\dag} )  \ .
\plabel{mellin}
\end{equation}
This yields the variation of $\zeta$ with respect to
$\varphi$ and $\psi$:
\begin{eqnarray}
\delta \zeta_{D\, D^{\dag}} (s)&&=\frac 1{\Gamma (s)} \int_0^\infty dt
\, t^{s-1} {\rm Tr}\sum_n \frac{(-t)^n}{n!}\left(
2\delta \psi (D\, D^{\dag})^n -2\delta \varphi (D^{\dag} D)^n \right)
\nonumber \\
&&=\frac 2{\Gamma (s)} \int_0^\infty dt
t^{s}\, {\rm Tr} \left( -2\delta \psi D\, D^{\dag} \exp(-tD\, D^{\dag} )
+2\delta \varphi D^{\dag} D\exp (-tD^{\dag} D) \right) \nonumber \\
&&=\frac {2 \Gamma(1+s)}{\Gamma (s)} {\rm Tr} \left(
-2\delta \psi D\, D^{\dag} (-tD\, D^{\dag} )^{-s-1}
+2\delta \varphi D^{\dag} D (-tD^{\dag} D)^{-s-1} \right)
\nonumber \\
&&=-2s
{\rm Tr}\, ((D\, D^{\dag} )^{-s}\delta \psi -
(D^{\dag} D)^{-s}\delta \varphi )
\plabel{zetaDD}
\end{eqnarray} 
Thus the introduction of $DD^{\dag}$ has provided a means to 
achieve multiplicative factors for the two variations -- at least 
in flat space, but this is sufficient for our purpose. 
By differentiating Eq.\ (\ref{zetaDD}) with respect to $s$
one arrives at
\begin{eqnarray}
\delta \zeta'_{D\, D^{\dag}} (0)&=&-2(\zeta (0\vert \delta \psi ,
D\, D^{\dag} )-\zeta (0\vert \delta \varphi , D^{\dag} D))
\nonumber \\
&=&-2(a_1 (\delta \psi ,
D\, D^{\dag} )-a_1 ( \delta \varphi , D^{\dag} D))\; .
\plabel{zprime}
\end{eqnarray}
To evaluate $a_1$ in the first term on the right hand side of 
Eq.\ (\ref{zprime}) 
we again use the method of \cite{Gilkey}. Introducing yet another 
type of differential operator in spinor space, 
we represent the operator
$D\, D^{\dag}$ as
\begin{eqnarray}
D\, D^{\dag} & = & -(\hat g^{\mu\nu}{\cal D}_\mu {\cal D}_\nu 
+E)\; , \nonumber \\
{\cal D}_\nu & = &\partial_\nu +
\psi_{,\nu} -\varphi_{,\nu} -\gamma^5 {\epsilon^\mu}_\nu 
\varphi_{,\mu}\; , \qquad \hat g^{\mu\nu}=e^{2(\psi 
-\varphi)}\eta^{\mu\nu}\; , \nonumber \\
E & = &\hat g^{\mu\nu}\; (\hat\nabla_\mu \hat\nabla_\nu 
\varphi)\;  , 
\plabel{Eomega}
\end{eqnarray}
and again use the result (\ref{a1}). The covariant derivatives 
$\hat \nabla_\mu$ refer to the present metric $\hat g_{\mu\nu}$. 
In a similar manner the 
second heat kernel 
 coefficient $a_1$ for the operator $D^{\dag} D$
is obtained by the replacement $\varphi\to -\psi$,
$\psi \to -\varphi$. From Eq.\ (\ref{zprime}) with $\Delta = 
\eta^{\mu\nu}\partial_\mu \partial_\nu$ 

\begin{equation}
\delta \zeta'_{D\, D^{\dag}} (0)=-\frac{1}{3\pi}
\int  d^2 x \sqrt{-\eta} \left[ \delta \psi (2\Delta \varphi + \Delta \psi )
+ \delta \varphi (2\Delta \psi + \Delta \varphi )\right]
\plabel{a1DD}
\end{equation}
follows. 
The curved space version of Eq.\ (\ref{a1DD}) 
can be obtained  by means of the identity $\sqrt{-\eta} \Delta = 
\sqrt{-g} \Box = \sqrt{-g}\, g^{\mu\nu}\, \nabla_\mu \nabla_\nu$. 
We have retained the determinant also for the flat metric in 
order to cover the case of light cone coordinates (\ref{7}) where 
$\eta = \det \eta \neq -1$.

Now all variations of the effective 
action $W (\rho, \phi, \psi)$ with
respect to all background fields can be summarized: 

\begin{eqnarray}
\frac{\delta W}{\delta \varphi} & =& -\frac{1}{12\pi}\,
\sqrt{-\eta} 
(6 \partial^\mu (\rho \partial_\mu \varphi ) + 2{\Delta} \rho
- 2\Delta\psi -\Delta \varphi )\; , \plabel{var1} \\
\frac{\delta W}{\delta \psi} & = & -\frac{1}{12\pi}\,
 \sqrt{-\eta} 
(\Delta \rho - 2\Delta \varphi - \Delta \psi )\; , \plabel{var2} \\
\frac{\delta W}{\delta \rho} & = & 
-\frac{1}{12\pi}\, \sqrt{-\eta} 
(- \Delta \rho - 3(\partial_\mu \varphi )^2 + 2\Delta \varphi
+ \Delta \psi )\; . \plabel{var3}
\end{eqnarray}
The solutions to Eqs.\ (\ref{var1})-(\ref{var3}) can be found by 
inspection : 

\begin{equation}
W = -\frac 1{24\pi}\int\, d^2 x\, \sqrt{-\eta} (-\rho \Delta \rho +
2\psi \Delta \rho -\psi \Delta \psi  - 6\rho (\partial_\mu \varphi )^2
+4 \varphi \Delta \rho -4\varphi \Delta \psi -\varphi \Delta 
\varphi )\; .
\plabel{zpr}
\end{equation} 
By the replacements $\sqrt{-\eta}\, \Delta = \sqrt{-g}\Box$ when 
acting on $\varphi$ or $\psi$, and 
$\sqrt{-\eta}\Delta \rho = -\frac{1}{2} \sqrt{-g} R$   
and some partial integrations the integrated effective action  
can be brought into covariant form:

\begin{eqnarray}
 W & = & -\frac{1}{24\pi}\int\, d^2 x \sqrt{-g}\; 
 \bigg[ -\frac{1}{4} R \Box^{-1} R 
 +3 (\nabla \varphi)^2 {\Box^{-1}} R  
 - 2 R (\psi + \varphi) + \nonumber \\
&& \quad\quad + (\nabla \psi)^2 + (\nabla\varphi)^2 + 
 4(\nabla^\mu \psi) (\nabla_\mu \varphi)  \bigg] + W(\mu, \mu')\; . 
\plabel{newP}
\end{eqnarray}
The first term in Eq.\ (\ref{newP}) represents the Polyakov action 
\cite{Polyakov} for minimal coupling ($\varphi = \psi = 0$) of the 
scalar fields. $\varphi (\phi)$ and $\psi (\phi)$ encode a 
general dilaton coupling of the scalars and of the 
dilaton-dependent measure, respectively. Thus Eq.\ (\ref{newP}) 
generalizes the Polyakov action 
for the case of non-minimal coupling to the dilaton
field.  The appearance of a new nonlocal term  
should be emphasized. A functional integral applied 
to a bounded region in space time always contains ambiguities 
with respect to eventual surface variables. In that case 
Eq.\ (\ref{newP}) may 
acquire further (here undetermined) contributions. The terms 
$W(\mu, \mu')$ depending on the renormalization points $\mu, 
\mu'$ will be discussed below. 

For SGR from $D$ dimensions the case $\varphi = \psi = \phi$ is 
of special interest:

\begin{equation}
W_{SRG} = \frac{1}{96\pi} \; \int\, d^2 x \sqrt{-g} \left[ R 
\, \Box^{-1} R  - 12 (\nabla \phi)^2 \Box^{-1} R + 12 \phi R 
- 24 (\nabla \phi)^2 \right] + W(\mu, \mu')\; . 
\plabel{WSRG}
\end{equation}
The second, nonlocal term was not present in the analogous formula 
for the full effective action in 
\cite{Mukhanov}.  The first three terms, however, appear in the 
``uncorrected'' effective action there. In the first ref.\ \cite{antiev} 
all four terms, but the last two with  different factors, can be 
found. 

Usually, the full effective action including conformally invariant
part is available as a  power series in a small parameter
\cite{Zel}. No such parameter exists for the BH background.
Therefore, the closed form of the action (\ref{WSRG}) is 
essential. Two previous famous examples where such a closed
form could be obtained were the Polyakov and WZNW actions. In those
cases the effective actions were completely defined by the
corresponding anomalies. For general dilaton theories in $D=2$ we 
have here a similar situation, because the ``dilaton anomaly'' 
Eqs.\ (\ref{vecf}), (\ref{var2}) ultimately  can be interpreted 
as carrying the information of part of the conformal anomaly 
belonging to some theory in $D$ dimensions. But, as will be seen 
in the next section, Eq.\ (\ref{WSRG}) cannot be the whole story.

\section{Hawking radiation}

As pointed out above, the direct derivation of the Hawking flux to 
${\cal J}^+$ from  Eq.\ (\ref{WSRG}) has to rely on a complete functional 
integration of the action. This is avoided in the 
Christensen-Fulling approach \cite{CF77} where only an ordinary 
integration is required. 

In conformal light cone coordinates Eqs.\ (\ref{7}), (\ref{8}) we 
separate the conformal anomaly (\ref{T3}) 
for $\varphi = \psi = \phi$ as

\begin{eqnarray}
T_{+-} & = & T^{min}_{+-} + T^{(1)}_{+-}\; , \plabel{45} \\
T^{min}_{+-} & = & - \frac{1}{12\pi} \partial_+ 
\partial_- \rho\; , \plabel{46} \\
T^{(1)}_{+-} & = & \frac{1}{4\pi} (\partial_+ \partial_- \phi - 
(\partial_+\phi)(\partial_-\phi))\; . \plabel{47} 
\end{eqnarray}
{}From Eqs.\ (\ref{var1}) and (\ref{var2}) we obtain 

\begin{equation}
\frac 1{\sqrt{-g}} \frac{\delta W}{\delta \phi}=
-\frac{1}{4\pi} (2\partial^\mu (\rho\partial_\mu \phi )
+\Delta \rho -2\Delta \phi )\; .
\plabel{deltaW}
\end{equation}
Equation (\ref{cons}) for the minus component of the index $\nu$
becomes

\begin{eqnarray}
\partial_+ T_{--} & = & - \partial_- T_{+-} + 2 
(\partial_- \rho)\, T_{+-} - \nonumber \\
&& \quad - \frac{(\partial_- \phi)}{2\pi} \left[ \partial_+ 
\partial_-\rho + \partial_+ (\rho \partial_- \phi) + 
\partial_-(\rho \partial_+ \phi) - 2 (\partial_+ 
\partial_-\phi)\right]\; . \plabel{49}
\end{eqnarray}

{}From Eqs.\ (\ref{7}) and (\ref{8}) the external fields only
depend on $z(u)$, therefore Eq.\ (\ref{49}) may be integrated
straightforwardly.  Choosing the limits $z_h = - \infty$ and $z
= \infty$ for $T_{--}$ we take into account the condition of a
finite flux (in Kruskal coordinates) at the horizon \cite{CF77}
which means vanishing $T_{--}$ at $z = - \infty $.  The
integrated first term on the right hand side of Eq.\ (\ref{49})
only contributes at the limits of the integral.  It vanishes
there (cf.\ Eqs.\ (\ref{6})-(\ref{8})).  The integral from
$T_{+-}^{min}$ in the second term on the right hand side of Eq.\
(\ref{49}) produces the flux for minimal coupling (cf.\ Eq.\
(\ref{9})):

\begin{equation}
T_{--}^{min}  =  \frac{(D-3)^2}{192\, \pi\, u_h^2} = 
\frac{\pi}{12} {T_H}^2\; .\plabel{50}
\end{equation}
If expressed in terms of the Hawking temperature it does not 
depend on the dimension. Of course, the corresponding flux of the 
unreduced theory acquires an additional factor proportional to  
${T_H}^{D-2}$ from proper counting of further degrees of freedom 
on $S^{D-2}$.  
Inserting the nonminimal contribution Eq.\ (\ref{47}) from the 
anomaly $T_{+-}^{(1)}$ 
into the same term of Eq.\ (\ref{49}) after integration yields 

\begin{eqnarray}
T_{--}^{(1)} & = & - 2 \; \int\limits_{-\infty}^\infty \; dz'\; 
(\partial_{z'}\rho) T_{+-}^{(1)} = - \frac{1}{8 \pi} 
\int\limits_{-\infty}^\infty dz' (\partial_{z'}\rho) \left[ 
(\partial_{z'}\phi)^2 - \partial_{z'}^2\phi) \right] = \nonumber \\
& = & -  \frac{9 (D-2)}{2 (D-1)} \; T_{--}^{min} \; . \plabel{51}
\end{eqnarray}

In terms of the parameter $a $ in the models of Eq.\
(\ref{ldil}) this result has been obtained already in
\cite{KLVmpla}.  Together with $T_{--}^{min}$ this would yield
the unphysical result of a negative flux.  However, the
nonconservation also implies the additional terms in the second
line of Eq.\ (\ref{49}).  The last, $\rho$-independent one only
contributes a total derivative to the $z$-integral which from
the explicit expression of $\phi$ in Eq.\ (\ref{5}) with Eq.\
(\ref{8}) again vanishes at the limits of integration.  The
remaining terms by partial integrals may be written as

\begin{equation}
T_{--}^{(2)} = - T_{--}^{(1)} + \frac{1}{8\pi} 
\int\limits_{-\infty}^\infty \; dz'\; \partial_{z'} \left[2 
(\partial_{z'} \phi) (\partial_{z'} \rho) + \rho (\partial_{z'} \phi)^2 
\right]\; , \plabel{52}
\end{equation}
where the second term on the right 
hand side of Eq.\ (\ref{52}) vanishes. We thus observe complete 
cancellation of the dilaton dependent terms in the flux. 

It should be noted that the non-conservation equation
(\ref{cons}) is nothing else than the $D$-dimensional
conservation condition for the energy--momentum tensor.  Hence
we are allowed to apply the Christensen--Fulling procedure
\cite{CF77} without changes.

We finally  compare this result to the one of a direct 
computation of $T_{--}\vert_{z \to \infty}$ from the functionally 
integrated effective action. 
The contribution from the last term in square brackets of Eq.\ 
(\ref{WSRG}) to the functional derivative with respect to 
$g_{\mu\nu}$ leads to a term proportional to $g_{\mu\nu}$ itself 
and to one proportional $(\partial_\mu\phi)(\partial_\nu\phi)$. 
For $T_{--}$ in conformal gauge $g_{--} = 0$ and 
$(\partial_-\phi)^2 \propto K^2/u^2$ vanishes at infinity
and at the horizon. 

In the framework of the zeta function regularization this 
is taken into account by the terms

\begin{equation}
W(\mu ,\mu')=\zeta_A(0)\log \mu +\frac 12 \zeta_{DD^{\dag}} (0)\log \mu' 
\plabel{mu}
\end{equation}
in $W_{SRG}$, Eq.\ (\ref{WSRG}).  Here we need two normalization
parameters, $\mu$ and $\mu'$, since we use zeta functions of two
different operators.  {}From $\zeta_A(0)=a_1(1,A)$ and the
explicit formulas (\ref{newA}), (\ref{a1}) and (\ref{Eomega}) it
can be verified that the only term in Eq.\ (\ref{mu}) which is
not a total derivative reads

\begin{equation}
\log \mu \int d^2 x \sqrt {-g} \left( -\frac 1{4\pi} 
(\nabla \phi )^2 \right)\; .
\plabel{only}
\end{equation}
However, the contribution of Eq.\ (\ref{only}) 
to $T_{--}$  vanishes at infinity as well as at the
horizon. 

The first three terms in Eq (\ref{WSRG}) coincide with the
integrated effective action of ref.  \cite{Mukhanov} before
there a compensating expression has been added.  Their
contribution to $T_{--}$ has been discussed recently in ref. 
\cite{Balbinot}.  Direct insertion of the background solution
(\ref{4})-(\ref{7}) corresponds to the choice of the so-called
Boulware vacuum \cite{CF77}.  It implies vanishing flux at
infinity for all those terms, including the one for minimal
coupling.  At the same time the conformal flux at the horizon is
finite which entails a divergent flux in global (Kruskal)
coordinates.  On the other hand, if - as in our
Christensen-Fulling approach - the Unruh vaccum is chosen
(vanishing conformal flux at the horizon), the total flux from
Eq.\ (\ref{WSRG}) becomes negative, because the negative dilaton
dependent contribution is larger than the ("correct") positive
one from the Polyakov action.

In this context we should recall that the derivation of the
radiation flux to infinity is known ot be a quite delicate
matter \cite{FrNo}.  Even without dilatons (as in the full $D=4$
theory) the choice of the asymptotics for the inverse
d'Alembertian has a decisive influence on the result.  The
direct integration of the EM conservation also is completely
insensitive to backscattering effects which appear when the
equations of motion for the scalar field are solved in a BH
background.  This backscattering is influenced strongly by the
inclusion of dilaton fields \cite{Mukhanov}.

Of course, by certain explicit assumptions which treat the
different terms in Eq.  (\ref{WSRG}) in a different manner, our
result (\ref{50}) could be obtained also from that equation; for
example different choices could be made for the asymptotic
behavior of the inverse d'Alembertian in the Polyakov term
(leading to $T_{--}^{ min}$) and in the first dilaton term
(leading to a vanishing 'Boulware' flux at infinity).  This
certainly does not seem satisfactory; it just underlines our
opinion that the effective action approach has a fundamental
weakness: it encodes a UV effect from quantum corrections, i.e. 
in coordinate space is certainly only correct locally.  This is
in agreement with the rules for functional differentiation which
in an expression like (\ref{WSRG}) require sufficiently strong
vanishing of the fields at the infinite boundaries of the
integration in order to be able to perform partial integrations
without surface contributions.  But the region where the flux is
needed here is precisely at that boundary.  There the functional
derivative with respect to the metric, the flux, should give a
nonvanishing result.  Also the metric itself does not vanish
there but becomes Minkowskian.  The main advantage of the
approach used in our paper is that the input from the one loop
quantum effects entered locally.  The subsequent integral from
the horizon to infinity is a trivial, well defined ordinary one.

We conclude this section by noting that in the presence of the
non--minimal coupling to the dilaton the very definitions of
various vacua must probably be changed. Due to the dilaton
ordinary plane waves are no longer solutions of the field
equations. Hence the arguments based on selection of positive
frequency modes fail. Again we do not need to worry about such 
issues in our approach.

\section{Conclusions}

With Eq.\ (\ref{newP}) of our present paper we are able to
present the -- to the best of our knowledge -- first complete
derivation of the local part of the effective one-loop action in 
$D = 2$ for a
general dilaton theory.  The nonminimal coupling to scalar
fields encoded by $\varphi (\phi)$ and the dilaton measure $\psi
(\phi)$ may be specialized to any given dilaton model.  This
expression, as well as the one of Eq.\ (\ref{WSRG}) with
$\varphi = \psi = \phi$ for spherically reduced gravity from $D$
dimensions beside the Polyakov term contains another nonlocal
contribution.  Our derivation consistently uses $\zeta$-function
regularization for {\em all} terms.  We are able to tie in the
functional derivative for the dilaton field with a kind of
integrability condition involving the 2d scale anomaly together
with a contribution which refers to a flat background. 
Integrating the nonconservation of the energy momentum tensor we
find that the Hawking radiation at infinity is identical to the
one for minimally coupled scalars.

The cancellation of dilaton dependent terms in the flux does not 
seem too surprising in view of the fact that this is true also in 
the $D=4$ calculation. Thus the input for a complete $D=2$ computation 
should therefore be the same one as from the $D=4$ anomaly. In 
$D=2$ the effect of that anomaly is separated into the 
information encoded in the $D=2$ anomaly plus another 
contribution which is expressible as a functional derivative of 
the effective action with respect to the dilaton field. Indeed, the 
relation of the latter quantity to the ``transversal'' part of 
the $D=4$ anomaly has been pointed out already in \cite{Mukhanov} 
and \cite{Balbinot}.

However, 
taking the functionally integrated effective action as 
starting point, we 
encounter the usual problems \cite{Mukhanov,Balbinot}
related to the derivation of a 
global result - flux at infinity - from a quantity in which only local 
quantum corrections are encoded.
 Our method to integrate completely the effective action, however,
seems to allow interesting aplications in other fields, upon which we hope
to be able to report soon.

\section*{Acknowledgments}

One of the authors (W.K.) thanks H.\ Balasin and V.\ Mukhanov
for stimulating discussions.  We thank R.~Bousso and R.~Balbinot
for correspondence.  This work has been supported by the Fonds
zur F\"orderung der wissenschaftlichen Forschung project
P-12.815-TPH.  D.V.\ is also grateful to the Alexander von
Humboldt foundation and to the Erwin Schr\"odinger International
Institute for Mathematical Physics for support.


\begin{thebibliography}{99}

\bibitem{thom84} P. Thomi, B. Isaak and P. Hajicek,
                Phys. Rev. D {\bf 30} , 1168 (1984);
                P. Hajicek, Phys. Rev. D {\bf 30}, 1178 (1984);
                 S.R. Lau, Class. Quant. Grav. {\bf 13}, 1541 (1996).

\bibitem{KS92}W. Kummer and D.J. Schwarz, Phys. Rev. D {\bf 45}, 
3628 (1992).

\bibitem{Gro92}H. Grosse, W. Kummer, P. Pre$\check{s}$najder and 
                 D.J. Schwarz, J. Math. Phys. {\bf 33} (11), 
                 3892 (1992); T. Kl\"osch and T. Strobl, Class. Quant. 
                 Grav. {\bf 13}, 965 (1996), and {\bf 14},  
                 1689 (1997); W. Kummer and P. Widerin, 
                 Phys.\ Rev.\ D {\bf 52}, 
                 6965 (1995); W. Kummer and G. Tieber, {\em 
                 Universal conservation law and modified Noether 
                 symmetry in 2d models of gravity}, TUW-98-16, 
                 hep-th/9807122, Phys. Rev. D (to be published).

\bibitem{KKL97}M.O. Katanaev, W. Kummer and H. Liebl, Phys. Rev. 
                D {\bf 53}, 5603 (1996); Nucl. Phys. B {\bf 486}, 
                353 (1997). 

\bibitem{KS92-2}W. Kummer and D.J. Schwarz, Nucl. Phys. B {\bf 382}, 
                        171 (1992); F. Haider and W. Kummer, 
                        Int. J. Mod. Phys. A {\bf 9}, 207 (1994);  
 P. Schaller and T. Strobl, 
        Class.\ Quant.\ Grav. {\bf 11}, 331 (1994);   
                 W. Kummer, H. Liebl and D.V. Vassilevich, 
                 Nucl.\ Phys. B {\bf 493},  491 (1997) and B {\bf 
                 513}, 723 (1998); W. Kummer, H. Liebl and D.V. 
                 Vassilevich, {\em Integrating geometry 
                 in general 2D dilaton gravity with matter}, 
                 TUW-98-19, hep-th/9809168.   

\bibitem{Man91} G. Mandal, A. Sengupta and S..R. Wadia, 
  Mod. Phys. Lett. A {\bf 6}, 
                1685  (1991); S. Elitzur, A. Forge and E. Rabinovici, 
                Nucl. Phys. B {\bf 359}, 581 (1991), ; E. Witten, Phys. 
                Rev. D{\bf 44}, 314 (1991);
                 C.~G. Callan, S.~B. Giddings, J.~A. Harvey, and A.~Strominger,
                 Phys.\ Rev.\ D, {\bf 45}, 1005 (1992); J. Russo, 
                 L. Susskind and L. Thorlacius, Phys. Lett. B 
                 {\bf 292}, 13 (1992). 

\bibitem{deAl92}S.P. deAlwis, Phys. Lett. B {\bf 289}, 
                 278 (1992); T. Banks, A. Dabholkar, M. 
                 Douglas and M. O' Laughlin, Phys. Rev. D{\bf 
                 45}, 3607 (1992); S.D. Odintsov and I.L. 
                 Shapiro, Phys. Lett. B {\bf 263}, 183 (1991). 
\bibitem{CF77}S.M. Christensen and S.A. Fulling, Phys Rev. D {\bf 
15}, 2083 (1997).
\bibitem{Mukhanov}
V. Mukhanov, A. Wipf and A. Zelnikov, Phys. Lett. B {\bf 332},
283 (1994).
\bibitem{Chiba}
T. Chiba and M. Siino, Mod. Phys. Lett. A {\bf 12}, 709, (1997).
\bibitem{c-anom}
R. Bousso and S. Hawking, Phys. Rev. D {\bf 56}, 7788 (1997);
S. Nojiri and S. Odintsov, Phys. Rev. D {\bf 57}, 2363 (1998);
Mod. Phys. Lett. A {\bf 12}, 2083 (1997); 
Phys. Rev. D {\bf 57}, 4847 (1998); 
M. Buric, V. Radovanovic and A. Mikovic, ``One-loop corrections
for Schwarzschild black hole via 2-D gravity'', gr-qc/9804083.
\bibitem{KLVmpla}
W. Kummer, H. Liebl and D.V. Vassilevich, Mod. Phys. Lett. 
A {\bf 12}, 2683 (1997).
\bibitem{antiev}
R. Bousso and S. Hawking, Phys. Rev. D {\bf 57}, 2436 (1998);
R. Bousso, Phys. Rev. D {\bf 58}, 083511 (1998); S. Nojiri and
S.D. Odintsov, ``Can quatum corrected BTZ black hole
anti-evaporate?'', gr-qc/9806034; ``Quantum (in)stability of 2-D
charged black holes due to back reaction of dilaton coupled
scalars'', hep-th/9806055.
\bibitem{Ich}
S. Ichinose, Phys. Rev. D {\bf 57}, 6224 (1998).
\bibitem{KLVprd}
W. Kummer, H. Liebl and D.V. Vassilevich, Phys. Rev. D {\bf 58},
108501 (1998).
\bibitem{Dowker}
J.S. Dowker, Class. Quantum Grav. {\bf 15}, 1881 (1998).
\bibitem{Balbinot}
R. Balbinot and A. Fabbri, Hawking radiation by effective two--dimensional
theories, hep-th/9807123.
\bibitem{Russo}
F.C. Lombardo, F.D. Mazzitelli and J.G. Russo, 
Energy-momentum tensor for scalar fields
coupled to the dilaton in two dimensions, gr-qc/9808048. 
\bibitem{Polyakov}
A.M. Polyakov, Phys. Lett. B {\bf 103}, 207 (1981).
\bibitem{KKL}
M.O.\ Katanaev, W.\ Kummer and H.\ Liebl, Nucl.\ 
Phys.\ B {\bf 486}, 353 (1997).
\bibitem{LVA}
H.\ Liebl, D.\ V.\ Vassilevich and S.\ Alexandrov, 
Class.\ Quantum Grav.\ {\bf 14}, 889 (1997).
\bibitem{Toms}K. Fujikawa, U. Lindstr\"om, N.K. Nielsen, M. 
Rocek and P. van Nieuwenhuizen, Phys. Rev. D {\bf 37}, 391 (1988); 
D.J. Toms, Phys. Rev. D {\bf 35}, 3796 (1987). 
\bibitem{zeta1}
J.S. Dowker and R. Critchley, Phys. Rev. D {\bf 13}, 3224 (1976).
\bibitem{zeta2}
S.W. Hawking, Commun. Math. Phys. {\bf 55}, 133 (1977).
\bibitem{Atiyah}
M.F. Atiyah, V.K. Patodi and I.M. Singer, Math. Proc. Camb. Phil. Soc.
{\bf 79}, 71 (1976)
\bibitem{EKP}
G.~Esposito, A.Y.~Kamenshchik and G.~Pollifrone,
{\it Euclidean Quantum Gravity on Manifolds with Boundary},
(Kluwer, Dordrecht, 1997).
\bibitem{Gilkey}
P.B. Gilkey, J. Diff. Geom. {\bf 10}, 601 (1975);
{\it Invariance Theory, the Heat Equation, and the
Atiyah-Singer Index Theorem} (CRC Press, Boca Raton, 1994).
\bibitem{Zel}
G.A. Vilkovisky, in Quantum Theory of Gravity, ed. S.M.Christensen
(Hilger, Bristol, 1984);
A.O. Barvinsky and G.A. Vilkovisky, Nucl. Phys. {\bf B 282}, 163 (1987);
{\bf B 333}, 471 (1990); 
Yu. Gusev and A. Zelnikov, Class. Quantum Grav. {\bf 15}, L13 (1998).
\bibitem{FrNo}
I.D. Novikov and V.P. Frolov, {\it Physics of Black Holes}
(Kluwer, Dordrecht, 1989).
\end{thebibliography}
\end{document}